\documentclass[aps,prl,twocolumn,groupedaddress]{revtex4}
\usepackage{graphicx}

\begin{document}


\title{Observation of a quenched moment of inertia\\in a rotating strongly interacting Fermi gas}


\author{Bason Clancy, Le Luo, and John E. Thomas}
\email[jet@phy.duke.edu]{}
\affiliation{Duke University, Department of Physics, Durham, North
Carolina, 27708, USA}



\date{\today}

\begin{abstract}
We make a model-independent measurement of the moment of inertia
of a rotating, expanding strongly-interacting Fermi gas. Quenching
of the moment of inertia is observed for energies both below and
above the superfluid transition. This shows that a strongly
interacting Fermi gas with angular momentum can support
irrotational flow in both the superfluid and collisional normal
fluid regimes.
\end{abstract}


\maketitle

Strongly interacting Fermi gases~\cite{OHaraScience} provide a
unique paradigm for exploring strongly interacting nearly perfect
fluids in nature, from high temperature superfluids and
superconductors to exotic normal fluids, such as the quark-gluon
plasma of the Big Bang~\cite{Heinz,Shuryak},  or low viscosity
quantum fields~\cite{Son}. Rotating superfluids require
irrotational flow, which quenches the moment of inertia or
produces a vortex lattice~\cite{LondonRot}. Recently, the
observation of vortices has directly demonstrated superfluidity in
a strongly interacting Fermi gas~\cite{KetterleVortices}. However,
the moment of inertia has not been measured and the rotational
properties have not been characterized in the normal regime, where
irrotational flow is not required.

We report a measurement of the moment of inertia $I$ of a strongly
interacting Fermi gas of $^6$Li atoms in both the superfluid and
normal fluid regimes, by releasing rotating clouds from a
cigar-shaped optical trap. In the superfluid regime, a rapid
increase in the angular velocity is observed as the cloud expands,
indicating a quenching of $I$ to values as low as 0.05 of the
rigid body value. However, quenching persists for energies far
above the superfluid transition, in contrast to previous
measurements spanning half a century, in
nuclei~\cite{Migdal,WintherNuclei}, liquid
helium~\cite{FairbanksHe} and Bose-Einstein condensates
(BEC's)~\cite{Foote,Inguscio}, where irrotational behavior was
attributed only to superfluidity. Our results demonstrate that a
strongly interacting Fermi gas with angular momentum can support
irrotational flow not only in the superfluid regime, but also in
the normal fluid regime, which we attribute to nearly perfect
collisional hydrodynamics.

Strongly interacting Fermi gases exhibit hydrodynamic behavior in
both the superfluid and normal fluid regimes. While superfluidity
explains the low temperature hydrodynamics~\cite{Kinast}, the
origin of nearly ideal flow in the normal fluid remains an open
question~\cite{BruunViscous}. The hydrodynamics of the gas has
been studied in expansion
dynamics~\cite{OHaraScience,SalomonExpInt,JinExpansion,MITvortexexpansion},
and in collective
modes~\cite{Kinast,KinastDampTemp,Bartenstein,GrimmPrecmeas}, but
is also predicted to dramatically affect the rotational
properties~\cite{StingariSlowFermi}, which have been investigated
only in vortices~\cite{KetterleVortices,MITvortexexpansion}.
Vortices demonstrate irrotational hydrodynamic flow, but only form
at temperatures well below the superfluid
transition~\cite{KetterleVortices}. In contrast, measurement of
the moment of inertia provides a  probe of the rotational
properties in both the normal and superfluid regimes.

In our experiments, a strongly interacting Fermi gas is prepared
using a 50:50 mixture of the two lowest hyperfine states of $^6$Li
atoms in an ultrastable CO$_2$ laser trap with a bias magnetic field
tuned to a broad Feshbach resonance at $B = 834$
G~\cite{BartensteinFeshbach}. At 834 G, the gas is cooled to quantum
degeneracy through forced evaporation by lowering the trap depth
$U$~\cite{OHaraScience}. Then $U$ is recompressed to
$U_0/k_B=100\,\mu$K, which is large compared to the energy per
particle of the gas.

At the final trap depth $U_0$, the measured oscillation
frequencies in the transverse directions are $\omega_x=2\pi\times
2354(4)$ Hz and $\omega_y=2\pi\times 1992(2)$ Hz, while the axial
frequency is $\omega_z=2\pi\times 71.1(.3)$ Hz, producing a
cigar-shaped trap with $\lambda=\omega_z/\omega_x=0.030$. The
total number of atoms at is typically $N = 1.3\times 10^5$. The
corresponding Fermi energy $E_F$ and Fermi temperature $T_F$ for
an ideal (noninteracting) harmonically trapped gas at the trap
center are $E_F=k_B T_F=\hbar\,\bar{\omega}(3N)^{1/3}$, where
$\bar{\omega}=(\omega_x\omega_y\omega_z)^{1/3}$. For our trap
conditions,  $T_F = 2.4\,\mu$K.

After evaporation and recompression, we typically achieve energies
of $E = 0.56\,E_F$, which is close to the ground state energy of $E
= 0.50\,E_F$~\cite{Entropy}.  To heat the sample, energy is added by
releasing and then recapturing the gas, after which the gas is
allowed to reach equilibrium for 0.5 s. The total energy $E$ of the
cloud is determined in the universal, strongly interacting regime
from the mean square axial (z) cloud size, using $E=3
m\omega_z^2\,\langle z^2\rangle$, where $m$ is the atom
mass~\cite{Entropy,ThomasUniversal}.

\begin{figure}[tb]
\includegraphics[width=3.5in]{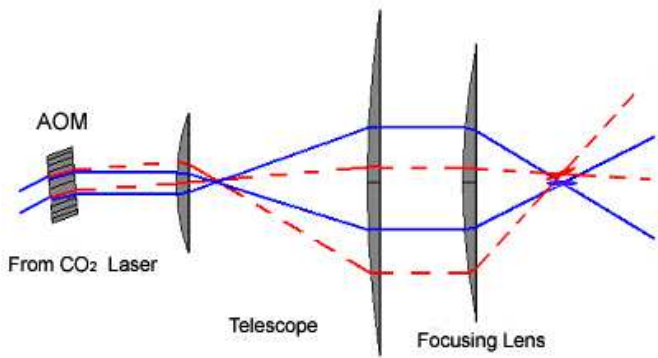}
\caption{Scheme to rotate the optical trap. The optical trap is
rotated by changing the frequency of an acoustooptic
modulator(AOM). The rotation adds angular momentum to the trapped
cloud before release. } \label{fig:fig1}
\end{figure}

Once the gas has been prepared in the desired energy state, it is
rotated by suddenly changing the direction of CO$_2$ laser beam as
shown in Fig.~\ref{fig:fig1}. Rotation of the cigar-shaped CO$_2$
laser trap is accomplished by changing the frequency of the
acousto-optic modulator (AOM) that controls the trap laser
intensity, using a radiofrequency (rf) switch.  When the frequency
is changed, the direction of the trapping laser beam changes,
causing the position of the beam on the final focusing lens to
translate. This translation causes primarily a rotation of the
cigar-shaped trap at the focal point, about an axis (y)
perpendicular to the plane of the cigar-shaped trap. A scissors
mode~\cite{StringariScissors} is excited by the rotation. Then the
cloud is permitted to oscillate in the trap for a chosen period
that determines the initial angular velocity of the cloud before
release.

\begin{figure*}[tb]
\includegraphics[width=7.0in]{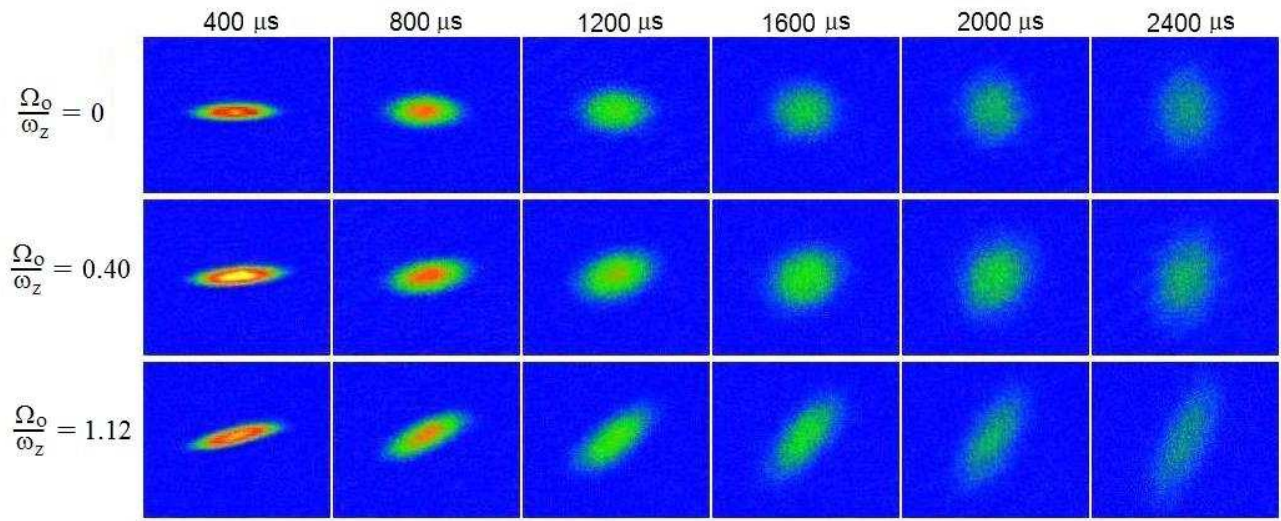}
\caption{Expansion of a rotating, strongly interacting Fermi gas.
The initial angular velocity  $\Omega_0$ is given in units of the
trap axial frequency $\omega_z$.} \label{fig:fig2}
\end{figure*}

Fig.~\ref{fig:fig2} shows cloud images as a function of expansion
time for the coldest samples, with a   typical energy $E = 0.56
E_F$ near the ground state. When the gas is released without
rotation of the trap, Fig.~\ref{fig:fig2} (top), the Fermi cloud
expands anisotropically, as previously predicted~\cite{Menotti}
and observed~\cite{OHaraScience,SalomonExpInt}. In that case, the
gas expands rapidly in the narrow (x,y) directions of the cigar,
while remaining nearly stationary in the long (z) direction,
inverting the aspect ratio $\sigma_x/\sigma_z$ as the cloud
becomes elliptical in shape.

Quite different expansion dynamics occurs when the cloud is rotating
prior to release, Fig.~\ref{fig:fig2} (middle) and (bottom), which
demonstrates irrotational flow in a nearly perfect hydrodynamic
regime. In this case, the aspect ratio $\sigma_x/\sigma_z$ initially
increases toward unity. However, as the aspect ratio approaches
unity, the moment of inertia decreases and the angular velocity of
the principal axes increases to conserve angular momentum as
previously predicted~\cite{StringariIrrot} and
observed~\cite{Foote,Inguscio} in a superfluid BEC. After the aspect
ratio reaches a maximum less than unity~\cite{StringariIrrot},  it
and the angular velocity begin to decrease as the angle of the cigar
shaped cloud approaches a maximum value less than 90$^\circ$.

\begin{figure*}[tb]
\includegraphics[height=2.5 in]{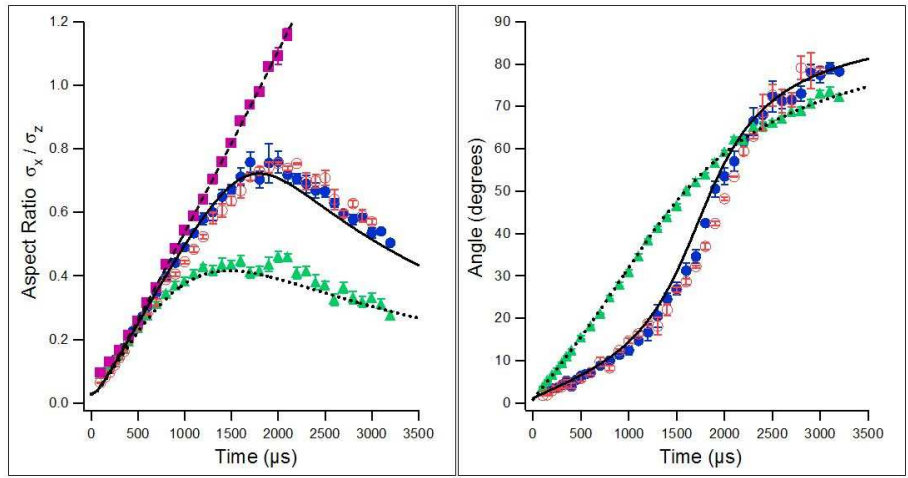}
\caption{Aspect ratio and angle of the principal axis versus time.
Purple squares (no angular velocity); Blue solid circles
($\Omega_0/\omega_z = 0.40$, $E/E_F = 0.56$); Red open circles
($\Omega_0/\omega_z = 0.40$, $E/E_F = 2.1$); Green triangles
($\Omega_0/\omega_z = 1.12$, $E/E_F = 0.56$). The solid, dashed,
and dotted lines are the theoretical calculations using the
measured initial conditions.} \label{fig:fig3}
\end{figure*}

Fig.~\ref{fig:fig3} shows the measured aspect ratio and the angle
of the principal axes versus expansion time, which are determined
from the cloud images. The measured density profiles are fit with
a two-dimensional gaussian distribution, which takes the form $A\,
exp[-a z^2- b z x-c x^2]$. From the values of a, b, and c, the
aspect ratio of the rotated cloud and the angle of the long axis
of the cloud with respect to the laboratory z-axis are determined.

We attempt to model the data for measurements near the ground
state (blue solid circles and green triangles of
Fig.~\ref{fig:fig3}), by using a zero temperature hydrodynamic
theory for the expansion of a rotating strongly-interacting Fermi
gas in the superfluid regime. A theory of this type was first used
to describe the rotation and expansion of a weakly interacting
BEC~\cite{StringariCritical,StringariIrrot}. The model consists of
the Euler and continuity equations for a superfluid, where the
velocity field $\mathbf{v}$ is irrotational, i.e., $\nabla \times
\mathbf{v}=0$. The driving force for the expansion arises from the
gradient of the chemical potential, which we take to be that of a
strongly interacting Fermi
gas~\cite{OHaraScience,StingariSlowFermi}. We also include the
force arising from magnet field curvature, which changes the
angular momentum at the point of maximum aspect ratio by 10$\%$
and the angle and aspect ratio at the longest release times by a
few percent. To determine the initial conditions for our model, we
directly measure the initial angular velocity and axial cloud
radius just after release, while assuming the transverse radii are
given by zero temperature values for our trap frequencies. The
results yield excellent agreement with all of the Fermi gas angle
and aspect ratio data, with no free parameters, as shown in
Fig.~\ref{fig:fig3}.

We make a model-independent measurement of the effective moment of
inertia $I\equiv L/\Omega$, where $\Omega$ is the angular velocity
of the principal axes of the cloud after release and $L=\Omega_0\,
I_0$ is the angular momentum, which is conserved during the
expansion (we neglect the small change arising from the magnetic
potential). The angular velocity $\Omega$ is calculated from the
time derivative of a polynomial fit to the angle versus time data.
To determine the initial moment of inertia $I_0$, we note that for
a cigar-shaped cloud with a small aspect ratio
$\sigma_x/\sigma_z$, the moment of inertia for the irrotational
fluid is nearly equal to the rigid body
value~\cite{Inguscio,StringariIrrot}. For our parameters
$I_0\simeq N m \langle z^2 \rangle_0$, within 0.3\% accuracy,
where $\langle z^2 \rangle_0$ is measured with respect to the
principal axes of the cloud. Correspondingly, the initial angular
momentum of the cloud is essentially equal to the rigid body value
just after release. The measured effective moment of inertia after
release is then $I=I_0\, \Omega_0/\Omega $. The corresponding
rigid body moment of inertia is determined from the fit to the
cloud profile, $I_{rig}=N m \langle x^2+z^2 \rangle$. Hence, we
obtain $I/I_{rig}=(\Omega_0/\Omega)I_0/I_{rig}$.

\begin{figure}[tb]
\includegraphics[width=3.5in]{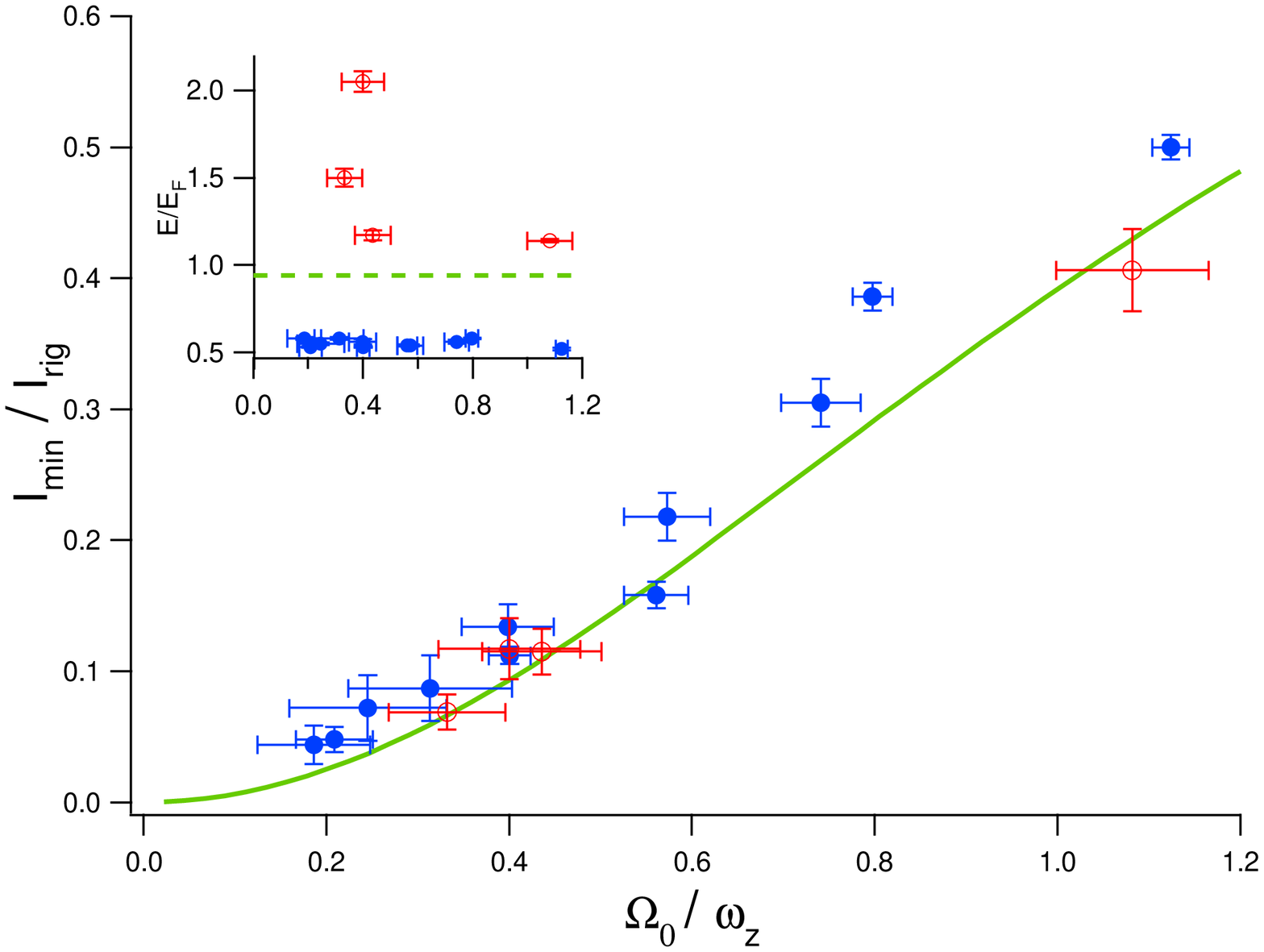}
\caption{Quenching of the moment of inertia versus initial angular
velocity $\Omega_0$. $I_{min}$ is the minimum moment of inertia
measured during expansion. $I_{rig}$ is the rigid body moment of
inertia corresponding to the cloud profile. Blue solid circles-
initial energy before rotation below the superfluid transition
energy $E_c =0.94\,E_F$. Red open circles- initial energy before
rotation above the superfluid transition energy. Insert shows the
energy for each data point. The dashed line shows the superfluid
transition energy $E_c$.} \label{fig:fig4}
\end{figure}

Fig.~\ref{fig:fig4} shows the measured minimum value of
$I/I_{rig}$ as a function of initial angular velocity $\Omega_0$.
The smallest values of $I/I_{rig}$ occur for the smallest
$\Omega_0$. For the coldest clouds (blue solid circles), where the
energy of the gas is close to that of the ground state, the gas is
believed to be in the superfluid
regime~\cite{KetterleVortices,Kinast,Entropy}. In this case, we
observe values of $I/I_{rig}$ as small as 0.05, smaller than those
obtained from the scissors mode of a BEC of
atoms~\cite{StringariMI,FootMI}. The solid line shows $I/I_{rig}$
as predicted by the superfluid hydrodynamic theory, which is in
very good agreement with the measurements.

Such nearly perfect irrotational flow usually arises only in the
superfluid regime. For example, normal weakly interacting Bose gases
expand ballistically above the critical temperature. We observe
ballistic expansion of the Fermi gas at 528 G, where the scattering
length vanishes.  In this case, after release of the rotating cloud,
there is no evidence of irrotational hydrodynamics. The aspect ratio
asymptotically approaches unity, and there is no increase in angular
velocity.

In contrast, for a normal strongly interacting Fermi gas, we
observe quenching of the moment of inertia. To investigate the
normal fluid regime, we increase $E$ above the transition energy,
which we estimate to be $E_c = 0.94\,E_F$~\cite{Entropy}. The open
red circles in Fig.~\ref{fig:fig3}, show the aspect ratio and
angle versus time for $E = 2.1\,E_F$ and an initial angular
velocity $\Omega_0/\omega_z=0.4$. The results are nearly identical
to those obtained for $\Omega_0/\omega_z=0.4$ in the superfluid
regime (blue solid circles). We attribute this result to nearly
perfect collisional hydrodynamics in the normal strongly
interacting fluid, although a complete many-body microscopic
description of this regime does not yet exist.

\begin{figure}[tb]
\includegraphics[height=2.5in]{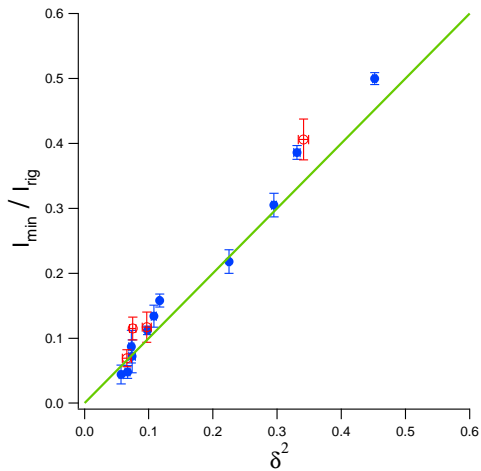}
\caption{Quenching of the moment of inertia versus the measured
cloud deformation factor $\delta^2$. Blue solid circles- initial
energy before rotation below the superfluid transition energy $E_c
=0.94\,E_F$. Red open circles - initial energy before rotation
above the superfluid transition energy.} \label{fig:fig5}
\end{figure}

We see from Fig.~\ref{fig:fig4}  that the moment of inertia is
quenched for energies both above and below the superfluid
transition. Irrotational hydrodynamics generally requires the
quenched moment of inertia to be given by~\cite{StringariIrrot}
\begin{equation}
I/I_{rig}=\delta^2\equiv\langle z^2-x^2\rangle^2/\langle
z^2+x^2\rangle^2,
 \label{eq:quenching}
\end{equation}
where $\delta^2$ is computed with respect to the principal axes.
Fig.~\ref{fig:fig5} compares the measured minimum values of
$I/I_{rigid}$ with the values of $\delta^2$ obtained from the
measured cloud aspect ratios, which directly verifies this
prediction.

We have observed quenching of the moment of inertia of a strongly
interacting Fermi gas over a wide range of energies, demonstrating
nearly perfect irrotational flow and low viscosity hydrodynamics
in both the superfluid and normal regimes. Our observations show
that these properties, which have been suggested as signatures of
superfluidity in the past~\cite{KersonTransport}, also appear in
the normal strongly interacting fluid.  These results have
important implications for other strongly interacting systems in
nature. It is known that quark-gluon plasmas, as created recently
in heavy ion accelerators, exhibit minimum viscosity hydrodynamics
and elliptic flow ~\cite{Heinz,Shuryak}. It is therefore possible
that irrotational flow will alter the signature of quark-gluon
plasmas with finite angular momentum.

This research was supported by the Chemical Sciences, Geosciences
and Biosciences Division of the Office of Basic Energy Sciences,
Office of Science, U. S. Department of Energy, the Physics
Divisions of the Army Research Office and the National Science
Foundation.


\end{document}